\documentclass[aps,pre,superscriptaddress,onecolumn]{revtex4}

\usepackage{amsfonts}
\usepackage{amsmath}
\usepackage{multirow}
\usepackage{longtable}
\usepackage{graphicx}
\usepackage{color}

\newcommand{\beginsupplement}{%
        \setcounter{table}{0}
        \renewcommand{\thetable}{S\arabic{table}}%
        \setcounter{figure}{0}
        \renewcommand{\thefigure}{S\arabic{figure}}%
     }

\beginsupplement

\begin{document}

\title{Modelling Opinion Dynamics in the Age of Algorithmic Personalisation}

\author{Nicola Perra}
\email[]{n.perra@greenwich.ac.uk}
\affiliation{Centre for Business Network Analysis, Business School, University of Greenwich, SE10 9LS, London, United Kingdom}

\author{Luis E C Rocha}
\email[]{luis.rocha@greenwich.ac.uk}
\affiliation{Centre for Business Network Analysis, Business School, University of Greenwich, SE10 9LS, London, United Kingdom}
\date{\today}

\begin{abstract}
Modern technology has drastically changed the way we interact and consume information. For example, online social platforms allow for seamless communication exchanges at an unprecedented scale. However, we are still bounded by cognitive and temporal constraints. Our attention is limited and extremely valuable. Algorithmic personalisation has become a standard approach to tackle the information overload problem. As result, the exposure to our friends' opinions and our perception about important issues might be distorted. However, the effects of algorithmic gatekeeping on our hyper-connected society are poorly understood. Here, we devise an opinion dynamics model where individuals are connected through a social network and adopt opinions as function of the view points they are exposed to. We apply various filtering algorithms that select the opinions shown to users i) at random ii) considering time ordering or iii) their current beliefs. Furthermore, we investigate the interplay between such mechanisms and crucial features of real networks. We found that algorithmic filtering might influence opinions' share and distributions, especially in case information is biased towards the current opinion of each user. These effects are reinforced in networks featuring topological and spatial correlations where echo chambers and polarisation emerge. Conversely, heterogeneity in connectivity patterns reduces such tendency. We consider also a scenario where one opinion, through nudging, is centrally pushed to all users. Interestingly, even minimal nudging is able to change the status quo moving it towards the desired view point. Our findings suggest that simple filtering algorithms might be powerful tools to regulate opinion dynamics taking place on social networks.
\end{abstract}

\maketitle

\section*{Introduction}

Various disciplines, as for example Sociology, Psychology and Behavioral Genetics, investigate the mechanisms leading to opinion formation in groups of people~\cite{Latane1981, Isenberg1986, Onofrio1999, Olson2001, Watts2007}. Generally speaking, opinions are formed by a combination of self-reflection, external information sources and real-world experiences that contribute to the individual reasoning process. Furthermore, it has been argued that social interactions are fundamental in this process since they can be used to spread and exchange information (e.g.\ word-of-mouth and communication) as well as to shape opinion formation (e.g.\ social pressure and social support) through peer's social influence. In this context, both the authority and number of contacts affect the opinion of individuals~\cite{Muchinik2013, castellano2009statistical, baronchelli2018emergence}. Apart from the natural process of influencing the opinions as well as behaviours of family, friends and co-workers~\cite{Christakis2007,Paluck2016}, social influence has been studied in marketing, politics and health interventions~\cite{Aral2012, Bond2012, Kim2015}.

The increasing popularity of social media augments the share on daily social interactions and thus their importance in the dynamics of information exchange, potentially enhancing the power of online social influence. In fact, the low cost of interacting with multiple people, simultaneously or asynchronously, without geographical constrains, facilitates the exchange of information at a faster pace and with a diverse set of contacts. Despite such unprecedented possibilities for interactions and consumption of information, we are still bounded by cognitive and temporal constraints~\cite{watts2002simple, falkinger2007attention, weng2012competition, wu2007novelty, leskovec2009meme, lerman2010information, ratkiewicz2010characterizing, onnela2010spontaneous, gonccalves2011modeling, dunbar1998social, dunbar2015structure,weng2018attention}. As a consequence, ideas, memes, individuals, companies and institutions compete for our limited attention which, in the current landscape, has acquired a real economic value~\cite{simon1971designing, davenport2001attention}. In this context, social platforms use algorithms tailored to improve the online experience by ordering and filtering information judged relevant to a particular individual (or social media user). Seemingly innocuous, algorithms are designed to keep users engaged and select which information is presented to them. They act as gatekeepers and intermediaries of information, functions that traditionally have been covered by news papers and thus by hand curated editorial choices~\cite{bozdag2013bias}. The shift towards algorithmic curation seems a natural consequence of the digital revolution. However, its short and long term societal effects are far from clear and matter of a heated debate. Before summarising some of the central points of such discussion, it is important to provide a better idea about the mechanisms driving algorithmic personalisation in the context of online social platforms. Indeed, algorithms are ruling a wealth of applications that go beyond social media. Examples are search engines, online shops and services.

Every time we access Facebook, or similar platforms, we experience algorithmic curation. In fact our \emph{News Feed} is the complex result of what people and companies connected to the Internet do. In $2013$ Facebook reported that every time we visit our News Feed we could be exposed on average to $1,500$ stories from our friends, people and pages we follow and other content that is sponsored~\cite{facebook_news, rader2015understanding}. Out of this pool, they prioritise around $300$ stories. While the details of the algorithm behind the personalisation are a corporate secret, we know that they are based on a combination of explicit information we provide such as demographics, likes, friendship relations, interests and implicit information derived from previous interactions with the platform, IP addresses, locations, cookies, posts age, page relationships, and content quality etc~\cite{bozdag2013bias, devito2017editors}. All this information is then fused with three main principles of content curation: popularity, semantic and collaborative filtering~\cite{moller2018not}. The first refers to the practice of promoting content that is popular across the platform. The second recommends posts similar to those previously consumed. Similarity is measured according to a range of features such as topics, words, authors, among others. The third refers to the tendency to suggest what people similar or close to us consume.

The critics of algorithmic personalisation consider such practices possibly very dangerous. Recommendations might increase the divide, exacerbating the phenomena of filter bubbles and echo-chambers~\cite{bozdag2015breaking, pariser2011filter, del2016echo}. They might remove opposing view points and homogenise the type of information we are exposed to. The phenomenon is often described as the \emph{threat of invisibility} because, in a society where more and more people use social platforms to gather information and form opinions, also what is not shown to us might be very important~\cite{bucher2012want, rader2017examining, ciampaglia2018algorithmic}. The homogenisation of information based on past preferences and actions might also reduce the possibility of serendipitous discoveries of new interests, sources, people, etc~\cite{devito2017editors}. Algorithmic gatekeeping might suffer from technical biases induced by the limitations of the databases they use and the codes they are based on~\cite{friedman1996bias, gillespie2014relevance}. Algorithms are data hungry. This raises ethical concerns about the access and sharing of potentially sensitive information~\cite{darren2015privacy, mittelstadt2016ethics}. Finally, social media platforms are large for-profit corporations driven by shareholders' and stakeholders' interests. There is clearly, at least in principle, a friction between societal and commercial objectives that might result in the so called \emph{corrupt personalisation}~\cite{darren2015privacy, sandvig2015can}. The proponents of algorithmic personalisation paint a far less scarier picture. First, without algorithmic filtering social platforms would be much more demanding and probably less appealing to users~\cite{darren2015privacy, liu2012personalization}. Indeed, as mentioned above, the data produced nowadays is simply too much for any type of manual filtering. Personalisation might be positive with respect to alternatives simply based on popularity metrics which could induce even stronger homogenisation of information~\cite{goldman2008search}. Some recent results suggest that the possible biases created by different personalisation algorithms are very similar to those resulting from hand curation~\cite{moller2018not}. Others show how the homogeneity and popularity bias are not the same across platforms and services~\cite{nikolov2018quantifying}. Finally, algorithmic gatekeeping can be developed having in mind the perils of filter bubbles and the importance of serendipity and heterogeneity in information~\cite{bozdag2015breaking}. To this point however, it is important to stress how the effects of algorithmic gatekeeping are largely not well understood. Thus the design of \emph{optimal} solutions are far from trivial and possibly counter intuitive~\cite{bail2018exposure}. Indeed, the complexity of the processes involved makes predicting the effects of new features very hard and the very idea of optimal is relative. In fact, at its roots, it depends also on the specific definition of democracy considered~\cite{bozdag2015breaking}.

In this background, we study, from a theoretical point of view, the effects of algorithmic personalisation on a population of networked individuals. We model online interactions where users both share their opinions in their timeline and update their own following the information exposed to them. In these settings, we consider the impact of different personalisation algorithms that prioritise the information shown to each user according to time ordering or their opinions expressed in previous interactions with the system. We then study a scenario in which a particular opinion is centrally posted in all users' timelines. We refer to this as \emph{nudge} and we assume that the social platform introduces, promotes, and pushes a particular opinion to all users thus nudging the population towards it. In all cases, we study the effect of each filtering method as a function of different i) initial conditions, and ii) network topologies where the interactions take place.

\section*{Materials and Methods}

We devise a mechanistic model to analyze the impact of social interactions and algorithmic exposure on the individual and group opinion dynamics. In particular, we consider a simple bipartisan model of opinion dynamics in which users can select between two opinions: $A$ or $B$. Users might decide to change their status according to the fraction of posts shown in their timelines promoting each opinion. Our model assumes that only part of the information shared by friends is shown to the users. This is the fundamental idea beyond algorithmic curation.  We then study how different methods used to create the personalisation affects the evolution of opinions in the system. We first look at the effect of sorting information randomly, according to time or previous ideological leanings expressed by each user in previous posts (semantic filtering). Then we consider a scenario in which one specific opinion is pushed in the timeline of each user. 

The opinion formation model contains three parts (Fig.~\ref{fig:01}). The first is the underlying social network structure connecting users through friendship ties. The second is the activation mechanism that defines the timings in which information is exchanged between users through the social network. The third part is the algorithmic filtering mechanism that selects which information is presented in the user's timeline (i.e.\ the posts actually seen by the user). 

The model has some similarities and is inspired by previous research. Wang et al~\cite{weng2012competition} proposed a model to mimic the spreading of memes on Twitter. In their approach, users' timelines are populated by posts from other users. With the aim to study the mechanisms behind the popularity of memes, they considered time ordered timelines in which posts survive only for a limited period of time. However, they did not considered opinion dynamics, but the spreading of memes. Also, they did not investigate the effects of different sorting algorithms. More recently, Rossi et al~\cite{rossi2018closed} shed some light, as we aim here, on the feedback between opinions and algorithmic personalisation. To this end, they proposed a model in which a single user interacts with a news aggregator which recommends news items to the user. In their model, they considered rather realistic scenarios in which the user might be subject to confirmation biases and the aggregator tries to optimise the number of clicks of the user. However, the model neglects the network effects induced by the interaction of multiple users connected via social ties. Finally, one of the areas that received more attention, probably due to data availability, focuses on the effects of algorithms on customer choices~\cite{bressan2016limits, fleder2009blockbuster, dandekar2013biased, spinelli2017closed}. Although, some of these papers study the interplay between algorithmic personalisation and social interactions, they consider purchase or more in general items selection (i.e.\ books) which are driven by different processes than those we are considering here. For example, purchasing power, popularity of items, reviews of products, among others, are critical ingredients in those cases.    

\subsection*{Network Structure}

The social network structure is fixed and defined as a set of $N$ users (i.e.\ network nodes) connected by undirected links $(i,j)$. The distribution of contacts, clustering, and small world phenomena are well known features of networks that might have a strong effects on opinion dynamics or more in general complex contagion processes~\cite{guilbeault2018complex, barrat2008dynamical, baronchelli2018emergence, barabasi2016network,centola2015spontaneous}. To isolate the role of topology and its interplay with the various algorithmic personalisation strategies, we consider three types of networks models. The first is a random network following the configuration model~\cite{newman2003structure} with power-law (heterogeneous) degree distribution ($P(k)\propto k^{-2.5})$ with $k_{\text{min}}=2$. These parameters produce small-world networks with an average degree $\langle k \rangle = 5.5$ and an average clustering coefficient $\langle C \rangle = 10^{-3}$.  The second is the Watts-Strogatz model~\cite{watts1998collective} where each node is initially connected to six (i.e.\ $\langle k \rangle = 6$) friends in a ring configuration; with probability $g$ a link is rewired to a uniformly chosen random node. In the following, we consider three configurations of the model by setting $g=0$, $g=10^{-2}$, and $g=1$. The three values of $g$ all produce networks with homogeneous degree distributions but allow us to explore very different combination of clustering and average path length. The first network is characterised by high values of clustering $\langle C \rangle = \frac{3(\langle k \rangle-2)}{4(\langle k \rangle-1)}=\langle C(0) \rangle=0.6$~\cite{barrat2000properties} and a very large average path length which is known to scale as $N/(2\langle k \rangle)$. The second instead, by a large value of the clustering $\langle C(g) \rangle =\langle C(0) \rangle (1-g)^{3}=0.58$~\cite{barrat2000properties}. However, in contrast with the first case, this network is characterised by the small-world phenomenon. The third, is instead a random network with negligible clustering ($\langle C \rangle=\frac{\langle k \rangle}{N-1}$) and characterised by the small-world phenomenon. The final type of network we consider is a regular $2D$ lattice with periodic boundary conditions. Each node has degree $4$, the clustering coefficient is zero, and the average path length is very large due to spatial clustering between nodes. In our simulations, we use $N=10^5$ and run the simulations taking averages over $10$ independent realisations of the networks with the same fixed parameters (except for the lattice which is not subject to random fluctuations). In summary (see Table~\ref{tab:01}), the first type of networks, CM for short, is characterised by a heterogeneous degree distribution, negligible clustering and short average path length. The second type of networks, WS for short, features homogeneous degree distributions as well as i) high clustering and large average path length (for $g=0$) ii) high clustering and short average path length (for $g=10^{-2}$) iii) negligible clustering and short average path length (for $g=1$). Finally, the regular lattice, LA for short, is characterised by a homogeneous degree distribution, zero clustering, and large average path length. Altogether, these networks allow to isolate and contrast the interplay and interactions of the three topological features with the sorting algorithms.

\begin{table}
\centering
\begin{tabular}{|l|r|r|c|}
\hline
\textbf{Network Type} & \textbf{Degree Distribution} & \textbf{High Clustering} & \textbf{Small World}  \\
\hline
CM & Heterogeneous  & x & \checkmark  \\
\hline
WS, $g=0$ & Homogeneous & \checkmark & x  \\
\hline
WS, $g=10^{-2}$ & Homogeneous& \checkmark &  \checkmark\\
\hline
WS, $g=1$ & Homogeneous & x  & \checkmark \\
\hline
LA & Homogeneous & x & x \\
\hline
\end{tabular}
\caption{Main networks' features summary. In the table CM stands for Configuration model, WS for Watts and Strogatz model, and LA for regular lattice.}
\label{tab:01}
\end{table}

\subsection*{Nodes Activation}

With probability $p_i$, each user $i$ becomes active during one time-step $t$. When the user becomes active, it first broadcasts its current opinion to all its friends, for example by writing a post, then updates its own looking at its personal timeline $R_i(t)$. In particular, a user whose timeline shows $30\%$ of posts promoting opinion $A$ and thus $70\%$ promoting opinion $B$, will adopt opinion $B$ with probability $0.7$ and opinion $A$ with probability $0.3$. In other words, the updates are done proportionally to the share of posts promoting each opinion in the timelines.  Although the most common opinion shown to each user will be most likely adopted, the process is not deterministic. In fact, there is a level of randomness which is an important component of opinion formation and dynamics~\cite{castellano2009statistical}. In the main text, we consider a scenario in which the activation probabilities are extracted from a power-law distribution ($F(p) \sim p^{-1.5}$ with $p \in [0.01,1]$). Indeed, analyses of real online social networks show that the propensity of each user to start a social interaction follows an heterogeneous distribution~\cite{perra2012activity, ribeiro2013quantifying}. In the Supplemental Information (SI), we consider the case of a constant activation probability ($p=0.1$, which is the average of the heterogeneous case) across the entire population. The results are qualitatively similar to the homogeneous case.

\subsection*{Algorithmic Personalisation and Filtering}

At time $t$, each user $i$ has a hidden list $L_i(t)$ containing all the opinions received from its friends since the last time $t_{\text{last}}$ that user $i$ was active, i.e.\ $L_i(t)$ contains all opinions posted by $i$'s friends during $[t_{\text{last}}, t)$, plus the opinions in user $i$ timeline at time $t_{\text{last}}$ ($R_i(t_{\text{last}})$). 

\begin{figure}[ht]
\centering
\includegraphics[scale=0.25]{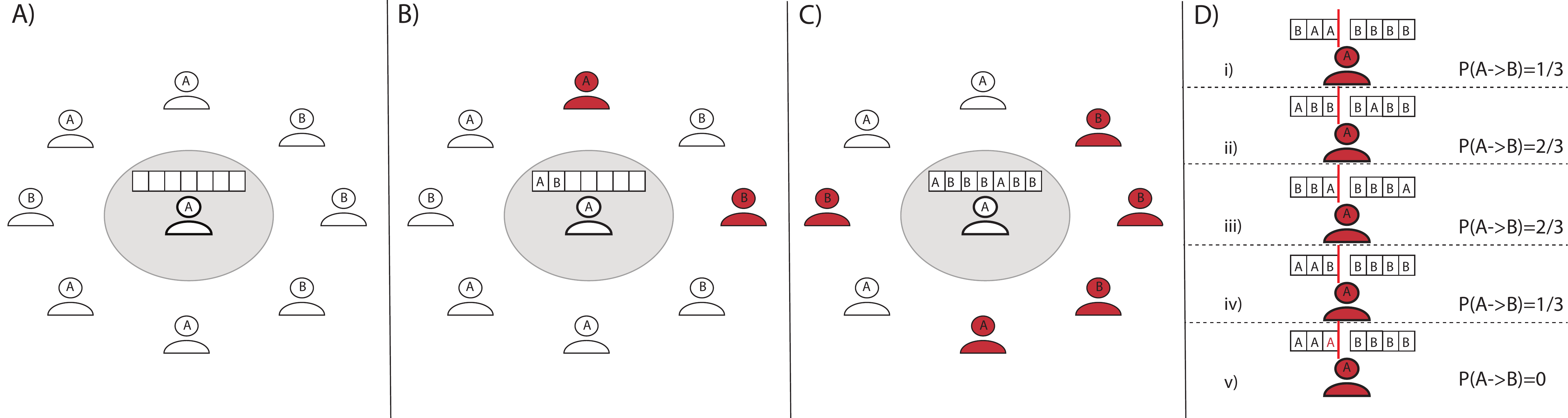}
\caption{Schematic representation of the opinion dynamics model. In the plot we focus on the user $i$ in the center of panels A-C. It is connected with eight other users that are depicted around. For simplicity, we assume that these users are connected only to the central user $i$ thus their opinions or timelines will not change or be updated until user $i$ will be active. Active users are represented in red and the list (i.e.\ $L_i$) of posts that user $i$ could possibly see is described by the boxes. Panel A shows the initial state. Opinions are drawn randomly and the $L_i$ is initially empty. At the next time step (panel B) two friends of $i$ are active and post their opinions thus $L_i$ starts to be filled. At the next time step (panel C) other friends become active and so on. In panel D we show what would happen in case user $i$ becomes active. We assume $Q=3$ (red vertical line) and show the effect of different algorithmic personalisation schemes: i) REF, the list is random shuffled; ii) OLD, does not change the order, the posts are shown in the order they were posted; iii) REC, recent posts are prioritized; iv) PR, shows the effects of a personalisation that prioritise posts according the preference/opinion of $i$; v) NU, shows the nudge case in which one post (in red), $z=1/3$, shown to the user does not come from its friends but it is imposed by the social platform. For each scenario we show the probability that the user $i$ would have to change opinion. In all scenarios only the first $Q$ posts after the personalisation affect the opinion of $i$. The rest are deleted.}
\label{fig:01}
\end{figure}

We consider different methods, see Figure~\ref{fig:01}D, to select items from the list $L_i(t)$ and thus create the timeline $R_i(t)$: (i) select $Q$ random items from $L_i(t)$ (reference method, named REF), (ii) show older posts first (old method, OLD), (iii) sort the list $L_i(t)$ from the most recent to the oldest received and show most recent posts first (recent method, REC), (iv) sort $L_i(t)$ according to $i$'s current opinion (e.g.\ if $i$'s opinion is $A$, sort the list as $AAA \ldots BBB$) (preference method, PR), (v) given any of the sorting method just described, a random fraction $z$ of opinions in this list is replaced by a pre-defined opinion centrally chosen (nudge method, NU). In all the scenarios the timeline $R_i(t)$ is a subset of this virtual list $L_i(t)$ and has maximum length $Q$. The remaining of the list is discarded once a timeline is formed at a given time $t$. In the main text we set $Q=20$, but as we show in the SI its value has little impact on the results.

\section*{Results}

In this section, we study the effects of sorting algorithms on the group and on the individual opinion dynamics by first looking at methods REF, OLD, REC and PR.  Then we investigate the effect of nudging, or centrally controlled bias. 

\subsection*{Opinion dynamics subject to algorithmic personalisation}

\begin{figure}[ht]
\centering
\includegraphics[width=\linewidth]{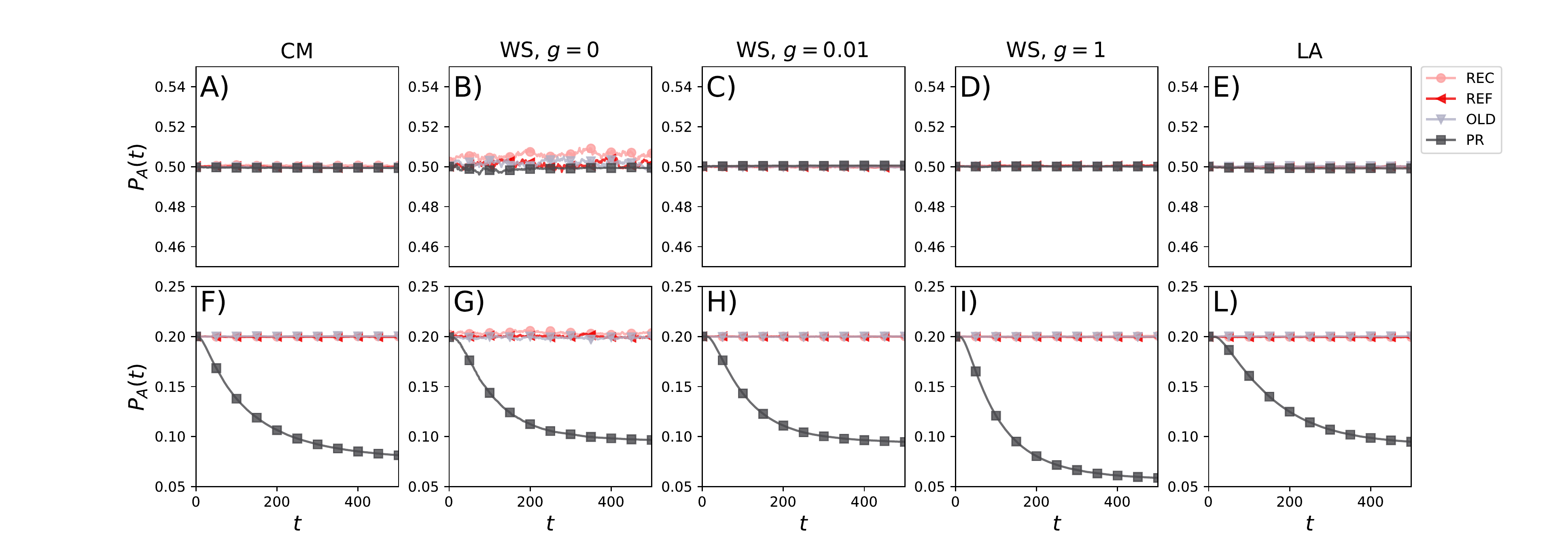}
\caption{Evolution of group opinion in the various network models. The prevalence $P_{\text{A}}(t)$ of opinion $A$ over time for starting $P_{\text{A}}(0)=0.5$ and $P_{\text{B}}(0) = 0.5$ (first row) and $P_{\text{A}}(0)=0.2$ and $P_{\text{B}}(0) = 0.8$ (second row). Each column describes the results for one of the networks. In each plot, we show results for the four sorting algorithms. Each plot is the average of $10^2$ independent simulations and to improve the visualisation we are showing the data points every $50$ time steps. In all scenarios we set $Q=20$.}
\label{fig:02}
\end{figure}

If we start the networks with two opinions $A$ and $B$, in equal proportions ($P_{\text{A}} = 0.5$ and $P_{\text{B}} = 0.5$) and uniformly distributed among the users (i.e.\ nodes), the prevalence of both opinions remains stable around the starting values for all four sorting methods for various network configurations (first row in Fig.~\ref{fig:02}). Thus, in these scenarios all the filtering algorithms are not able to break the status quo. Both ideas are able to coexist and the social platform does not drive the system towards a different equilibrium with respect to the initial status. If the opinions are unequal from the start ($P_{\text{A}} = 0.2$ and $P_{\text{B}} = 0.8$), the prevalence also remains stable for the methods REC, REF and OLD during the entire observation period (second row in Fig.~\ref{fig:02}). Indeed, as the order in which posts appears in the list $L_i$ of each user is random and regulated by the activation process of its neighbours, introducing a temporal bias (in the case of REC and OLD) does not affect the share ($20-80$) of opinions in each timeline thus preserving the status quo. On the other hand, the preference method (PR) leads to a decrease in the prevalence of the minority opinion ($P_{\text{A}}$) that mainly occurs within the initial $200$ time steps across all networks. 
In an unbalanced situation, the ordering of opinions according to users' preferences, for users holding opinion $B$, might lead to the disappearance of the opposite idea from their timeline. Effectively, such semantic filtering brings to zero the probability of such users to change idea (the \emph{visibility} of the others is zero) thus modifying the initial share and increasing the already dominant position of opinion $B$. It is important to notice that although the less popular opinion decreases its share, it is still able to survive. As shown in the SI, the survival happens even in case of much skewed initial configurations such as $1-99$. Results are similar for all network configurations, but for the WS network with $g=1$ the decrease in $P_{\text{A}}$ occurs earlier in time and $P_{\text{A}}$ stabilises at lower values ($P_{\text{A}}(500)= 0.06$, see Fig.~\ref{fig:02}-I). The CM network reaches larger values ($P_{\text{A}}(500)=0.08$) which are however slightly smaller than the other three networks ($P_{\text{A}}(500)=0.09$). Altogether, these results point to the fact that the PR filtering mechanism, in case of unbalanced initial conditions, causes more biases in networks characterised by null clustering, short path length, and homogeneous degree distributions. The heterogeneity in contact patterns hampers the reduction of the less popular opinion due to the presence of hubs which limit the effectiveness of the algorithm to hide the subordinate opinion.  High values of clustering both in combination with low and high average path length behave as in the case of regular lattice which features null clustering, high spatial correlations and high average path length. This result suggests that correlations in the connectivity patterns hamper the effectiveness of semantic filtering with respect to uncorrelated cases, which is qualitatively in line with previous research on recommender systems mentioned above~\cite{bressan2016limits}.   

\begin{figure}[ht]
\centering
\includegraphics[width=\linewidth]{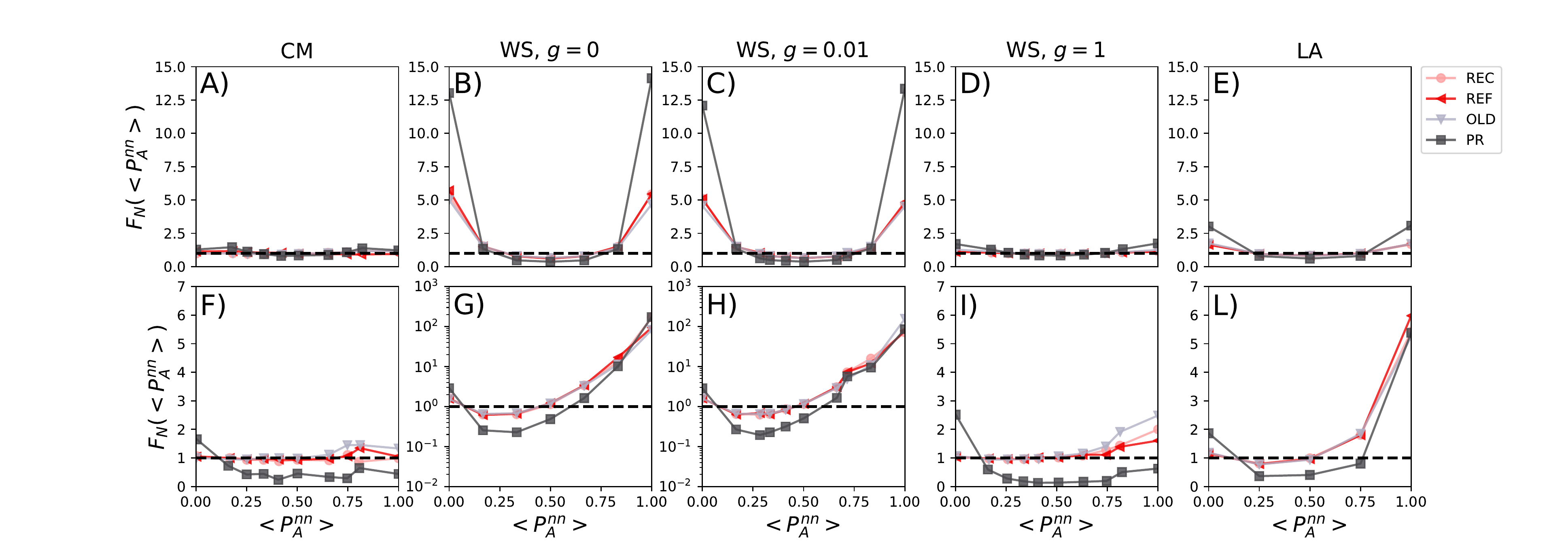}
\caption{Opinions of friends. We show the distribution of the fraction of friends (nearest neighbours, nn) of $i$ with the same opinion $A$ at $t=500$. We normalise the $y$-axis by dividing for the same quantity computed at $t=0$. In the first row we show the results for starting conditions $P_{\text{A}}(0)=0.5$ and $P_{\text{B}}(0) = 0.5$. In the second row, for starting conditions $P_{\text{A}}(0)=0.2$ and $P_{\text{B}}(0) = 0.8$. Each column describes the results for a particular network. Each plot is the average of $10^2$ independent simulations. In each plot, we consider the four ranking algorithms and set $Q=20$.}
\label{fig:03}
\end{figure}

The interplay between the features of the sorting mechanisms and the network structure might induce the formation of opinion clusters (echo chambers) and polarization in the population. Figure~\ref{fig:03} shows the distribution of the fraction of users $i$ whose friends (nearest neighbours,\emph{nn} for short) share the opinion $A$ at $t=500$, $\langle P_A^{nn} \rangle$. In these plots, the region of the $x$-axis close to one describes neighbourhoods formed by a 
majority of users sharing opinion $A$. Conversely, the region of the $x$-axis close to zeros describes neighbourhoods in which none, or very few users, adopt opinion $A$. In order to remove spurious effects, on the $y$-axis we show $F_N(\langle P_A^{nn} \rangle)$ that is obtained dividing the average distribution of opinions in each neighbourhood at time $t=500$ with its correspondent value at $t=0$. In doing so,  to avoid issues with the binning, we imposed $10$ equispaced bins ($\Delta x=0.1$) in all cases. 

For equal starting conditions ($P_{\text{A}} = 0.5$ and $P_{\text{B}} = 0.5$) (first row in Fig.~\ref{fig:03}), we notice that CM and WS ($g=1$) behave similarly. The distribution of opinions in the neighbourhood of all nodes is similar to the initial (random) configuration. The PR mechanism, in the case of WS with $g=1$, produces two little spikes at zero and one which indicate the emergence of low levels of polarisation in the network. Thus, the semantic filtering, in networks characterised by homogeneous degree distributions, null clustering, and short average path length, is able to produce few patches like-minded people which however do not form in case of heterogeneous connectivity patterns (CM network). The polarisation effects of the filtering algorithms, in particular PR, are much stronger in the other networks. Indeed, in highly clustered networks, independently of the average path length (WS graphs with $g=0$ and $g=10^{-2}$) the sorting mechanisms induce the emergence of strong opinion polarisation. As shown in the SI, if the size of the timelines is comparable with the degree (i.e. $Q=5$), REC and REF filtering methods produce stronger polarization effects in combination with high clustering in the contacts patterns. This is due to the fact that the PR mechanism in these cases might over represent the local minority opinion, thus hampering the convergence to a local majority. Echo chambers appear, although to smaller degree, also in case of strong spatial correlations and null clustering (LA network). These observations provide further evidence that topological as well as spatial correlations are important features for algorithmic filtering and that heterogeneity in contact patterns has a slightly stronger potential to hamper the formation of each chambers in comparison to homogeneous patterns.  In fact, in combination with the filtering mechanisms, these features induce a re-organisation of the distribution of opinions creating \emph{compact domains} made of like-minded nodes (echo chambers). This observation aligns with previous results in the literature of complex contagion processes on complex networks~\cite{dall2006nonequilibrium}. The polarisation emerges even if the total fraction of the population with opinion $A$ or $B$ is the same and stable as function of time (Fig.~\ref{fig:02}). Thus, although the interplay between the topological features and algorithm filtering does not change the status quo, it might introduce high levels of polarisation and echo chambering.

The case of unbalanced starting conditions (i.e.\ $P_{\text{A}} = 0.2$ and $P_{\text{B}} = 0.8$) confirms this picture (second row in Fig.~\ref{fig:03}). In fact, the behaviour of CM and WS ($g=1$) networks is still very similar.  The absence of topological clustering and the presence of short cuts between parts of the network (small-world phenomenon) does not allow the emergence of echo chambers around the dominant opinion. This is somehow surprising considering that opinion $B$ reaches around $90\%$ of the population. There is an exception however, the semantic filtering (PR method) fights against this tendency. As mentioned above, this personalisation algorithm reduces the \emph{visibility} of the few nodes holding opinion $A$ that might be present in neighbourhoods of $B$ nodes. The effective reduction of the share of nodes with opinion $A$ in such configuration brings the system to a different, and lower, equilibrium as eventually such nodes change opinion to $B$. This is confirmed by the fact that across the entire $x$-axis (with exception of $x=0$) the average number of neighbours holding some fraction of opinion $A$ decreases for all network models. The presence of heterogeneous connectivity patterns in CM networks reduces the effectiveness of such effect. Figure~\ref{fig:03}F-L  shows that the three other sorting algorithms, OLD above all, induce a relatively higher number of echo chambers around the subordinate opinion in WS ($g=1$) networks. As such effects are reduced in CM networks, we can speculate that this is due to the interaction between the algorithms and homogeneous contact patterns.

Also for unbalanced initial conditions, topological and spatial correlations induce a much stronger polarisation (notice the log scale on the $y$-axis). In particular,  in networks with high values of clustering, the number of $A$ echo chambers is two orders of magnitude larger than in the initial configuration. Spatial correlations introduce less stronger effects which are however of the order of a factor $6$. Overall, these results suggest that correlations interacts with the sorting algorithms creating compact domains that, in this case, boost the survival of the less common opinion with respect to the other two networks. 

\subsection*{Nudging Scenario}

\begin{figure}[ht]
\centering
\includegraphics[width=\linewidth]{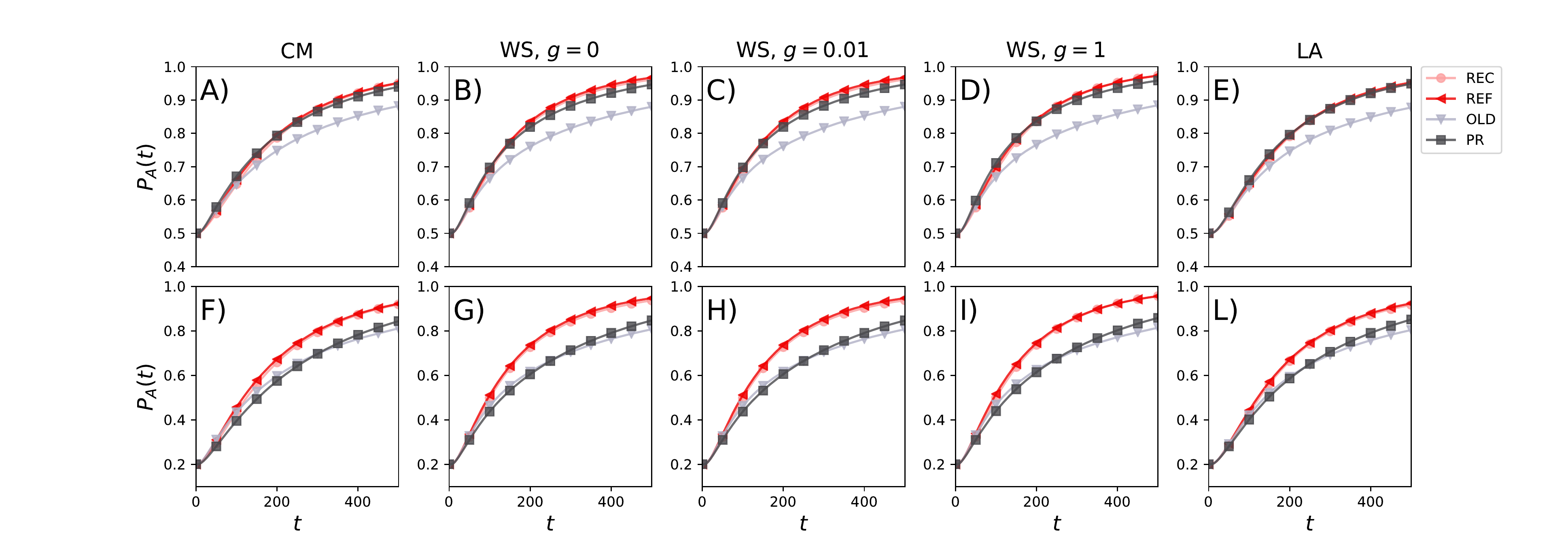}
\caption{Evolution of population opinion with central influencer (nudging). The prevalence $P_{\text{A}}(t)$ of opinion $A$ over time for starting conditions $P_{\text{A}}(0)=0.5$ and $P_{\text{B}}(0) = 0.5$ in the first row; $P_{\text{A}}(0)=0.2$ and $P_{\text{B}}(0) = 0.8$ in the second row. Each column describes the results for a particular network. In each plot, we set $Q=20$ and show results for the four sorting algorithms. Each plot is the average of $10^2$ independent simulations and to improve the visualisation data points are shown every $50$ time steps.}
\label{fig:04}
\end{figure}

We now consider the same filter algorithms studied in the previous section with the added feature that a random fraction $z$ of opinions, in the list $L_i(t)$ shown to each user $i$, is replaced by a pre-defined opinion centrally chosen and fixed in time (see Fig.\ref{fig:01}-D). In other words, we are investigating a scenario in which the social platform, where the interactions take place, nudges the users towards a particular opinion. In doing so, a fraction of the opinions shown to a user is substituted with ideas that might or might not be representative of what its nearest neighbours are posting. In the following, we set $z=0.1$, $Q=20$ and chose $A$ as the centrally set opinion. Independently of the starting conditions (balanced or unbalanced prevalence of $A$ and $B$), the group opinion moves towards $A$ for all network configurations and sorting algorithms (Fig.~\ref{fig:04}). In the case of unbalanced starting conditions (i.e.\ $P_{\text{A}} = 0.2$ and $P_{\text{B}} = 0.8$), opinion $A$ may become dominant as early as $t \sim 100$ time steps and typically no later than $t \sim 150$. At $t = 500$ almost the entire population turned to opinion $A$. Across the board, the REF method is the most efficient at $t=500$ to nudge the population opinion towards $A$. This is more evident with unbalanced initial conditions. Arguably, this is due to the fact that this filtering method (REF) is the least biased towards the current and past opinions hold by each user and its neighbours. The OLD method instead, in the balanced case, is significantly slower than the rest. By nature, this sorting algorithm slows down the overall drift of the system by showing to each user old neighbours' beliefs. In contrast to the results in the previous section, method PR performs in between the other methods.

\begin{figure}[ht]
\centering
\includegraphics[width=\linewidth]{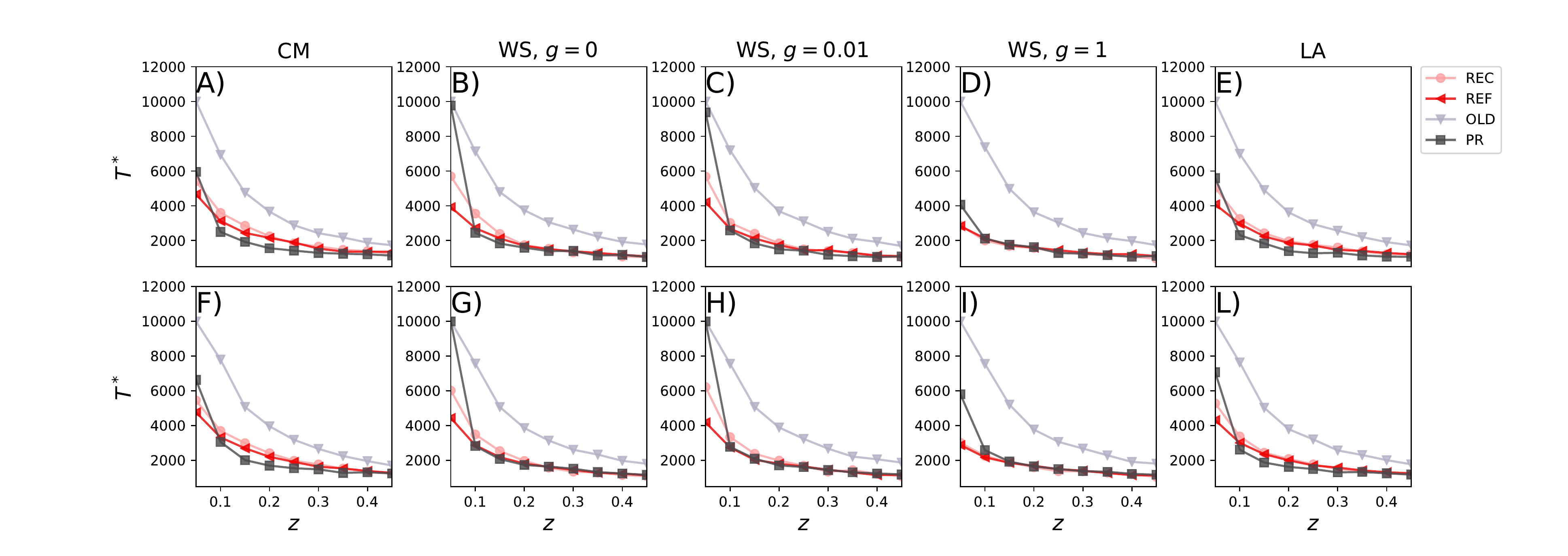}
\caption{Time for convergence. We show the time $T^*$ that the nudged opinion needs to converge (i.e.\ be shared by all nodes) as function of $z$. We set $10^4$ as maximal time of convergence of the simulations. In he first row we show the results for $P_{\text{A}}(0)=0.5$ and $P_{\text{B}}(0) = 0.5$; in the second row we show the results for $P_{\text{A}}(0)=0.2$ and $P_{\text{B}}(0) = 0.8$. In each plot, we set $Q=20$ and each column describes the results for a particular network model. Each plot is the average of $10$ independent simulations.}
\label{fig:06}
\end{figure}

In order to gather a better understanding of the impact of different sorting algorithms and networks topologies in the nudging process, in Fig.~\ref{fig:06} we show the time $T^*$ that the opinion $A$ needs to take over the entire population. To reduce the computational cost of the simulations, we set $10^{4}$ as time limit, thus the actual $T^*$ for the curves that reach such threshold is higher. The first observation is that the OLD sorting mechanism is, across the board, the slowest to reach convergence.  The OLD method introduces a delay by promoting the oldest posts of each neighbours, which, over time, might sponsor a different opinion with respect to the opinion the focus nodes currently hold. The second observation is that for $z=0.05$, thus when only one opinion is centrally replaced in the list of each user (since $Q=20$), the PR sorting algorithm is generally faster than the OLD method but slower than the others. This is particularly visible in the case of networks with topological (WS networks with $g=0$ and $g=10^{-2}$) or spatial correlations (LA network). As mentioned above, correlations interact with the semantic filtering (PR) creating relatively strong echo chambers which shield its members protecting them from the nudged opinion. The third observation is that for $z \ge 0.15$, the time for convergence for the REF, REC, and PR methods becomes a very weak function of $z$ and of the initial share of the two opinions. Fourth, as also noted in Fig.~\ref{fig:04}, the REF filtering mechanism, which does not introduce any type of bias in the ordering of neighbours' posts, is the fastest to reach convergence in all topologies. In the case of networks with negligible clustering and small average shortest path (WS networks with $g=1$) the convergence, for $z=0.05$, is reached
faster than in the other topologies. The comparison with the CM networks points one more time to the fact that heterogeneous connectivity patterns hamper both local (as seen before) and global convergence (as seen now) to a single opinion.

\begin{figure}[ht]
\centering
\includegraphics[width=\linewidth]{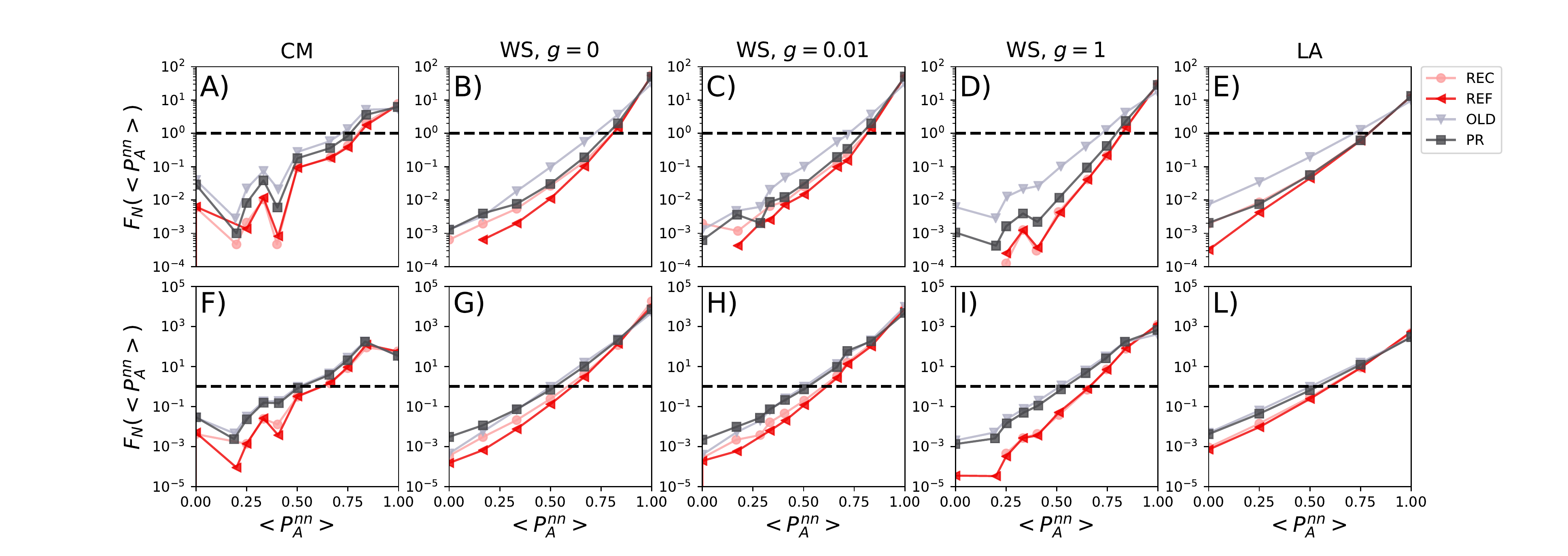}
\caption{Opinions of friends. The distribution of the fraction of friends of $i$ with the same opinion as $i$ at $t=500$. The $y$-axis is normalised by dividing by the same quantity computed at $t=0$. In the first row, we show the results for starting conditions $P_{\text{A}}(0)=0.5$ and $P_{\text{B}}(0) = 0.5$. In the second row, we show the results for starting conditions $P_{\text{A}}(0)=0.2$ and $P_{\text{B}}(0) = 0.8$. Columns show the results for different networks. Each plot is the average of $10^2$ independent simulations and we set $Q=20$.}
\label{fig:05}
\end{figure}

Figure~\ref{fig:05} shows the normalised distribution of the fraction of neighbours sharing opinion $A$ at $t=500$ for both starting conditions. Since opinion $B$ is almost extinguished in both cases, the distributions are all asymmetric and skewed towards values of $\langle P_A^{nn} \rangle$ close to one. Across all networks, the system is converging towards opinion $A$ and a large number of neighbourhoods share all the same nudged opinion. Indeed, the reduction of all the other values of $\langle P_A^{nn} \rangle$ is dramatic (note the logarithm scale on the $y$). The comparison between the CM and WS ($g=1$) networks shows how the heterogeneity in connectivity patterns alone hampers the formation of echo chambers around the dominant opinion.  

\section*{Discussion}

Technology has dramatically reduced the cost of communications and vastly increased their reach. As consequence, the amount of information produced at any given moment far surpasses our ability to consume it. Individuals, companies, and institutions alike compete for our attention that has acquired a real economic value. For example, online social platforms such as Twitter and Facebook profit from our engagement with their systems. Thus, optimising our experience is a key part of their business model. To this end, they adopt algorithmic personalisation which broadly describes a range of automated methods that filter the information to which we are exposed to. The filtering is not random, but targeted to stimulate our interests and online activity. Generally speaking the aims of algorithmic filtering are benign. However, in our hyper-connected society they might have unintended and unanticipated consequences. One area of concern is the possibility that such methods, which often are black boxes, affect opinion formation and dynamics. As many of our social interactions are mediated by such platforms and since much of the information gathering and exposure happen online, the perception of our friends as well as of specific issues might be distorted by such algorithms. These concerns are now mainstream and the words \emph{echo chamber}, \emph{confirmation bias}, and \emph{filter bubbles} have become a staple in the news cycle. Despite the heated debate the effect of algorithmic filtering on opinion dynamics is still largely not understood. 

In this paper, we studied the effect of different filtering algorithms on the opinion dynamics of a population of interconnected individuals. We modelled their interactions having in mind prototypical online platforms where users post their opinions which appear in friends' timelines. Due to the limited attention of each user, timelines are curated by means of algorithmic personalisation. We considered different scenarios where posts are selected i) randomly, ii) as function of the time when they were posted, and iii) as function of previous posts of users (semantic filtering). Even more, we studied how the features of the connectivity patterns mediating the interactions between users interplay with filtering algorithms and possibly affect opinions. In particular, we considered three different types of networks which allow us to isolate and contrast three main properties of real networks i) heterogeneity in degree distributions, ii) high values of clustering, and iii) small-world phenomenon. Finally, we investigated another scenario in which the social platform where interactions take place tries to nudge the population towards a particular opinion by manipulating a fraction of the posts in users' timelines.

For simplicity we investigated a bipartisan system in which users can adopt one of two opinions, i.e.\ $A$ or $B$. We studied two very different initial configurations where the share of opinion $A$ was equal ($50-50$) or much lower ($20-80$) than $B$. In these settings, we found that, independently of the social network structure, the sorting mechanisms are not able to change the status quo when the prevalence of opinions is equally distributed in the population. Despite the overall share of opinions does not change, we observed that some features of the networks interact with the filtering algorithms inducing polarisation effects. In particular, we found that topological (high values of clustering) and to a lesser extent spatial (proximity of nodes in a regular lattice) correlations create compact domains (echo chambers) formed only by like-minded individuals. Despite the algorithmic personalisation, absence of correlations, especially in conjunction with heterogeneity in the contacts patterns, hamper the formation of such echo chambers. In case of unbalanced initial conditions, we found that semantic filtering induces a further increase of the predominant opinion because of positive reinforcement.  Also in this scenario correlations play a significant role hampering the reinforcement by creating compact domains around the subordinate opinion that effectively is able to protect such minorities. As mentioned above the absence of topological, or the presence of spatial, correlations hinders the formation of echo chambers, even those centred around the dominant opinion. 

If a centralised bias mechanism (i.e.\ nudging) is introduced, population opinion can move towards the nudged opinion relatively fast and eventually  completely switch the status quo. In this case, filtering algorithms that randomly select friends' opinions or that give preference to the most recent ones speedup convergence to the centrally desired opinion. Conversely, methods biased towards older posts are the slowest to reach convergence. 
Semantic filtering, in case only a single post in users timelines is manipulated, slows down convergence especially in networks characterised by topological or spatial correlations. In these cases, the emergence of echo chambers opposes the change towards the nudged opinion.

Overall, our results support the view that opinion dynamics can be manipulated by algorithmic personalisation methods. Furthermore, they highlight the interplay between filtering mechanisms and the features of the networks where interactions take place. In particular, correlations in contacts patterns are key ingredients that might lead to the emergence of polarisation and echo chambers even in case of equal share between opinions. Our results show that filtering algorithms might exacerbate polarisation but the organisation of social ties is a key factor.  Conversely, the absence of correlations, especially in conjunction with heterogeneous distributions of contacts hampers the creation of compact domains. These results are in line with the literature of complex contagion processes on complex networks~\cite{baronchelli2018emergence, guilbeault2018complex, barrat2008dynamical}. The study of the nudging scenario shows how the dominant  opinion in the population can completely switch within short time by moderately pushing a desired opinion. This idea is supported by the Nudge Theory in behavioural economics originally developed by Richard Thaler~\cite{Thaler2008} and by recent experiments on the emergence of conventions~\cite{centola2018experimental}. The vulnerability of social media to manipulations can be disastrous because nudging for example can be easily introduced by the company controlling the media, by external players through paid advertisement or by the use of bots~\cite{ferrara2016rise}. There are claims that such strategies have been used to bias political public opinion in recent years~\cite{Bessi2016, Stukal2017}. Not least, since positive reinforcement mechanisms may increase opinion polarisation according to our study, the spread of fake news, that tend to spread faster than true news~\cite{Vosoughi2018,qiu2017limited}, may generate local convergence of opinion that later leads to the emergence of bubbles and group isolation~\cite{Nikolov2015}. In light of democratic access to information, a common and intuitive idea is that social media developers should make efforts to minimise opinion biases by increasing the diversity of opinions received from friends and counter-balancing the content of advertisements. However, recent empirical evidence points out that the actual dynamics at play are much more complex~\cite{bail2018exposure}. Indeed, the exposure to opposite view points might further increase polarisation~\cite{bail2018exposure}. Thus, the research on the feedback between opinion dynamics and algorithmic personalisation is just in its infancy.

%Model limitations and future directions
Our model aims to capture key mechanisms driving online social interactions and opinion dynamics. It has however some limitations since it disregards other potentially relevant mechanisms. In particular, our model does not consider that constantly diverging friends may stop following each other and thus their opinions are not shared. This manual curation, complementary to the automated curation included in our model, might further contribute to the formation of opinion bubbles and likely boosts biases towards the users' own opinion. Future models should include manual curation taking into account the user's own memory of his or her friends opinions. In the language of modern Network Science this will imply the development of adaptive and time-varying networks models that couple the dynamics of the networks (link formation) with the dynamics on the networks (opinion dynamics). Another limitation is that users have some level of resistance to change and thus some people need reinforced exposure to an opinion before switching. This may affect the timing, possibly delaying, and the likelihood to switch the prevalent opinion in the population. Also, friends do not necessarily have the same authority and possibly information coming from specific friends, e.g.\ close offline friends or relatives, may have higher importance on shaping the user's opinions. Another limitation is the type of filtering mechanisms considered. Future work should consider more complex and realistic models possibly linked to the history and features of each user as well as the popularity of specific memes or information circulating in the system. Finally, here we considered only some prototypical networks models as way to isolate three of the main features of real networks. Other properties such as the presence of communities, high-order correlations, and temporal connectivity patterns will be matter of future explorations. 

\bibliography{sample}

\section*{Acknowledgements}

The authors thank Dr. Andrea Baronchelli for his interesting suggestions and insights.

\section*{Author contributions statement}

N.P. and L.R. conceived the model, N.P. conducted the experiments, N.P. and L.R. analysed the results, wrote the manuscript and approved the manuscript. 

\section*{Additional information}

\textbf{Competing Interests}: The authors declare that they have no competing interests.

\newpage

\section*{Supplemental Information for ``Modeling opinion dynamics in the age of algorithmic polarization''.}

Here, we provide supplemental information about the algorithmic filtering methods, the opinion dynamics model, and we present a sensitivity analysis to study the robustness of the results to the variation of the main parameters.

\subsection{Algorithmic Personalization Mechanisms}

Due the limited attention of each node, we assume that the social platform, where interactions take place, shows to each user a maximum  of $Q$ posts. Consider a node $i$. It is characterized by an activity $p_i$ which determine the rate at which it engages with the social platform. This is extracted from a distribution $F(p)$. In the main, text we considered an heterogenous distribution, in section~\ref{sensitivity} we consider the case of a constant activity across the whole network. Suppose that at time $t$ node $i$ holds opinion $A$ which it was adopted by looking at the posts contained in its timeline, $R_i(t_{last})$, the last time ($t_{last}$) it was active. Suppose that the node is now active. It first broadcasts its opinion to all its neighbors (which might or might not see it due the filtering mechanism). Between time $t_{last}$ and $t$ some of its neighbors might have been active and broadcasts their opinions. These go to an hidden list $L_i(t)$ which contains also the posts in $R_i(t_{last})$. In case the list $L_i(t)$ is larger than $Q$, some posts are not shown to the user and deleted. In fact the user will just see posts in its timeline $R_i(t)$. The sorting process is defined by the algorithm personalization mechanism under consideration. We considered four possible approaches:
\begin{itemize}
\item posts are selected randomly. This is the reference method (REF for short)
\item posts are ordered considering the time when they were posted and the latest $Q$ are selected. We call this method REC, that stands for recent
\item posts are ordered considering the time when they were posted and the oldest $Q$ are selected. We call this method OLD, as older posts are selected first
\item posts are ordered considering the current opinion of the user which was formed by looking at the posts in its timeline and was the subject of a post at time $t$. This mimic semantic filtering methods as the filtering preferentially show to the user similar posts respect to what it recently broadcasted. We call this method PR for short. 
\end{itemize}
Each method selects $Q$ posts from the list $L_i(t)$, thus in general $|R_i(t)|\le Q$.

\subsection{Opinion Dynamics Model}

After the algorithmic filtering, the node $i$ is presented with a curated timeline $R_i(t)$. Remember that we considered a bipartisan system in which nodes can adopt one of two opinions, i.e. $A$ or $B$. Thus the timeline, generally, contains $a|R_i(t)|$ posts in favor of $A$ and $b|R_i(t)|$ in favor of $B$. The user will then adopt opinion $A$ with probability $a$ and opinion $B$ with probability $b$. Thus, as mentioned in the main text the adoption process, although biased towards the majority, is not deterministic. As long $a$ and/or $b$ are greater than zero the user can adopt the two opinions.   

\subsection{Sensitivity Analysis}
\label{sensitivity}

Below, we will show the results for i) constant activation rates ii) different $Q$ sizes iii) more unbalanced initial conditions.

\subsubsection{Constant activation rates}

\begin{figure}[ht]
\centering
\includegraphics[width=\linewidth]{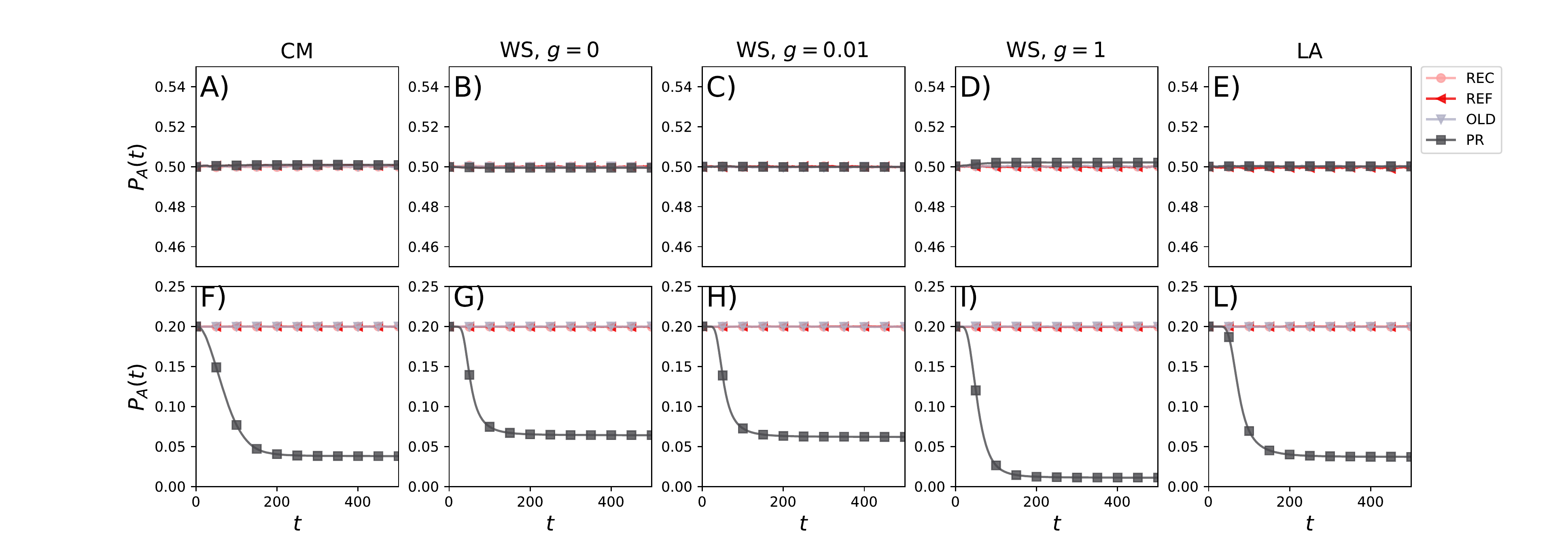}
\caption{Evolution of group opinion in the various network models. The prevalence $P_{\text{A}}(t)$ of opinion $A$ over time for starting from $P_{\text{A}}(0)=0.5$ and $P_{\text{B}}(0) = 0.5$ (first row) and $P_{\text{A}}(0)=0.2$ and $P_{\text{B}}(0) = 0.8$ (second row). Each column describes the results for one of the networks.  In each plot, we show results for the four sorting algorithms. Each plot is the average of $10^2$ independent simulations and to improve the visualization we are showing the data points every $50$ time steps. In all scenarios we set $Q=20$ and assume that each node as a constant activation probability $p=0.1$.}
\label{fig:SI01}
\end{figure}

In the main text, we considered heterogenous activity patterns following a distribution $F(p)\sim p^{-1.5}$ with $p \in [0.01,1]$. In these settings, the average activity $\langle p \rangle = 0.1$. In this section, we show the results considering that each node has the same activity $p=0.1$. In Figure~\ref{fig:SI01}, we show the behavior of $P_A$ as function of time for all the network topologies. In the first row, we show the results for a balanced initial condition in which $P_A(0)=P_B(0)=0.5$, in the second row instead $P_A(0)=0.2$ and $P_B(0)=0.8$. The results are in qualitatively in line with what shown in the main text: none of the sorting mechanisms is able to break the status quo in case of balance initial conditions. Instead in the case of unbalanced initial start the PR mechanism reduces the fraction of the subordinate opinion. Furthermore, the largest reduction happens in the case WS networks with $g=1$. Conversely, high values of the clustering coefficient, independently of the average shortest path, hamper the reduction as well as heterogenous connectivity patterns.

\begin{figure}[ht]
\centering
\includegraphics[width=\linewidth]{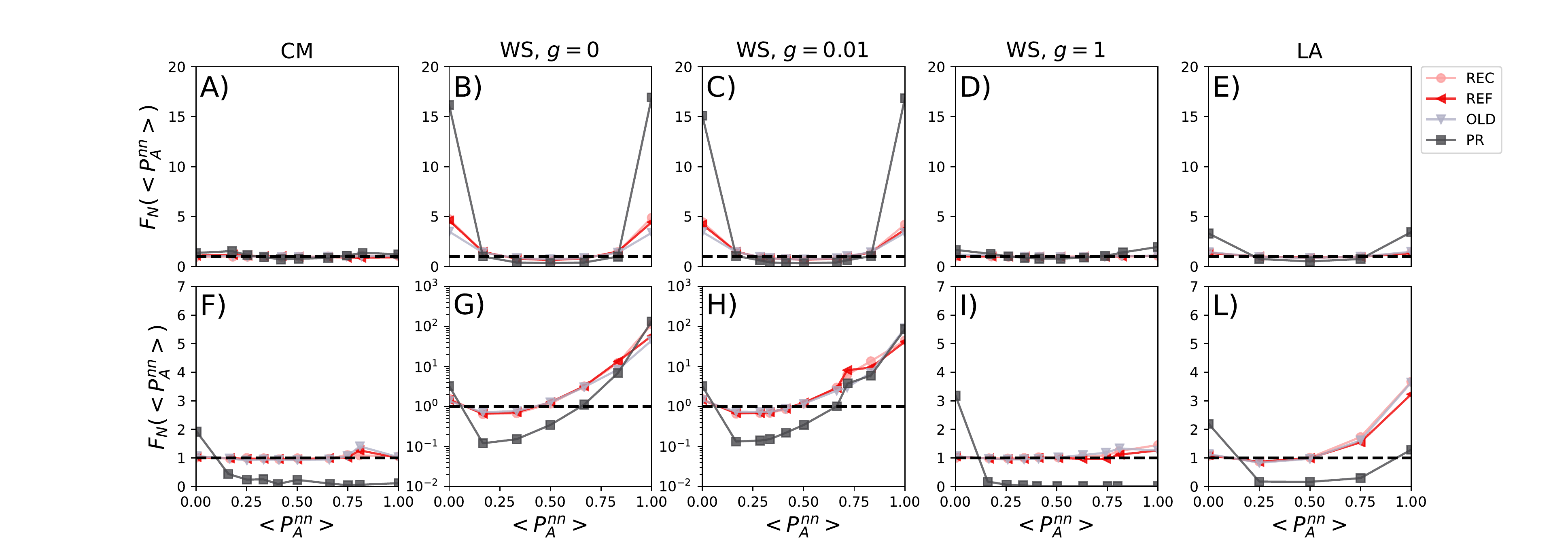}
\caption{Opinions of friends. We show the distribution of the fraction of friends (nearest neighbors, nn) of $i$ with the same opinion $A$ at $t=500$. We normalize the $y$-axis by dividing for the same quantity computed at $t=0$. In the first row we show the results for starting conditions $P_{\text{A}}(0)=0.5$ and $P_{\text{B}}(0) = 0.5$. In the second row, for starting conditions $P_{\text{A}}(0)=0.2$ and $P_{\text{B}}(0) = 0.8$. Each column described the results for a particular network. Each plot is the average of $10^2$ independent simulations. In each plot, we consider the four ranking algorithms and set $Q=20$ and assume that each node as a constant activation probability $p=0.1$.}
\label{fig:SI02}
\end{figure}

In Figure~\ref{fig:SI02}, we show the distribution of the fraction of neighbours holding opinion $A$ $\langle P_A^{nn} \rangle$. In order to highlight the difference respect to the initial conditions, on the $y$-axis we divided the distribution values at $t=500$ for the same quantity computed at $t=0$, which is then normalized by the initial conditions. The results are in line with what shown in the main text for heterogenous activity patterns. In particular, for a balanced starting condition (first row in the figure) both topological (WS with $g=0$ and $g=0.01$) induce the emergence of polarization effects. These are particularly visible for the semantic filtering (PR). Absence of clustering and heterogeneity in connectivity patterns are not conductive for such phenomena to emerge. For unbalanced initial conditions (second row in the Figure), the effects of clustering are much stronger. Semantic filtering induces the formations of echo chambers around the dominant opinion across all network topologies. However, as shown in the main text, heterogeneity in the connectivity patterns hamper this tendency.

\subsection{Different queue sizes}

\begin{figure}[ht]
\centering
\includegraphics[width=\linewidth]{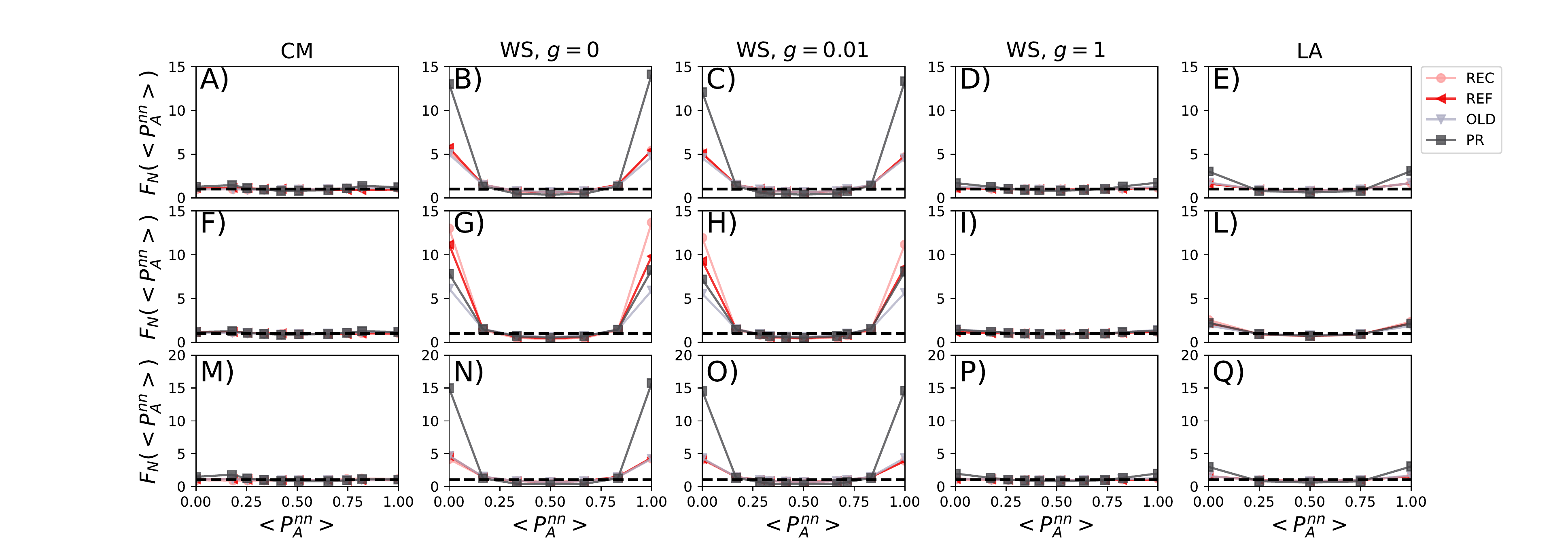}
\caption{Opinions of friends. We show the distribution of the fraction of friends (nearest neighbors, nn) of $i$ with the same opinion $A$ at $t=500$. We normalize the $y$-axis by dividing for the same quantity computed at $t=0$. We show the results for starting conditions $P_{\text{A}}(0)=0.5$ and $P_{\text{B}}(0) = 0.5$. In the first row, we show the results for $Q=20$, in the second for $Q=5$, in the third for $Q=50$. Each column described the results for a particular network. Each plot is the average of $10^2$ independent simulations.}
\label{fig:SI03}
\end{figure}

We now turn to study the role of the queue sizes $Q$. In Figure~\ref{fig:SI03} we show, for initial balanced conditions, the normalized distribution of the fraction of neighbors holding opinion $A$ for $Q=20$ (first row), $Q=5$ (second row), and $Q=50$ (third row). The results are qualitatively similar for all the values of $Q$. There are some differences in the case of small queue sizes ($Q=5$) in case of high clustered topologies (WS networks with $g=0$ and $g=0.01$). Here, despite all the sorting mechanisms give rise to polarization, their efficiency is a bit different than for larger values of $Q$. Indeed, for $Q=5$ the methods REF and REC, that provide either a random sample or a the latest posts, produce larger values of polarization than the PR method. Intuitively this is due to the fact that in case $Q$ is small than the average degree of nodes (which in for WS networks is $6$) the PR mechanism might reduce the polarization effects. To understand this effect think about the extreme situation in which nodes have all $Q=1$. Consider a node $i$ connected with $5$ nodes having opinion $A$ and only one with opinion $B$. Assume that $i$ holds opinion $B$. In case $Q=1$ has long as there is at least one neighbor with the same opinion the node will not change to $A$. Instead for larger values of $Q$, despite the bias, the opinion $A$ is likely to be more represented and eventually lead to a formation of cluster around that opinion.  

\begin{figure}[ht]
\centering
\includegraphics[width=\linewidth]{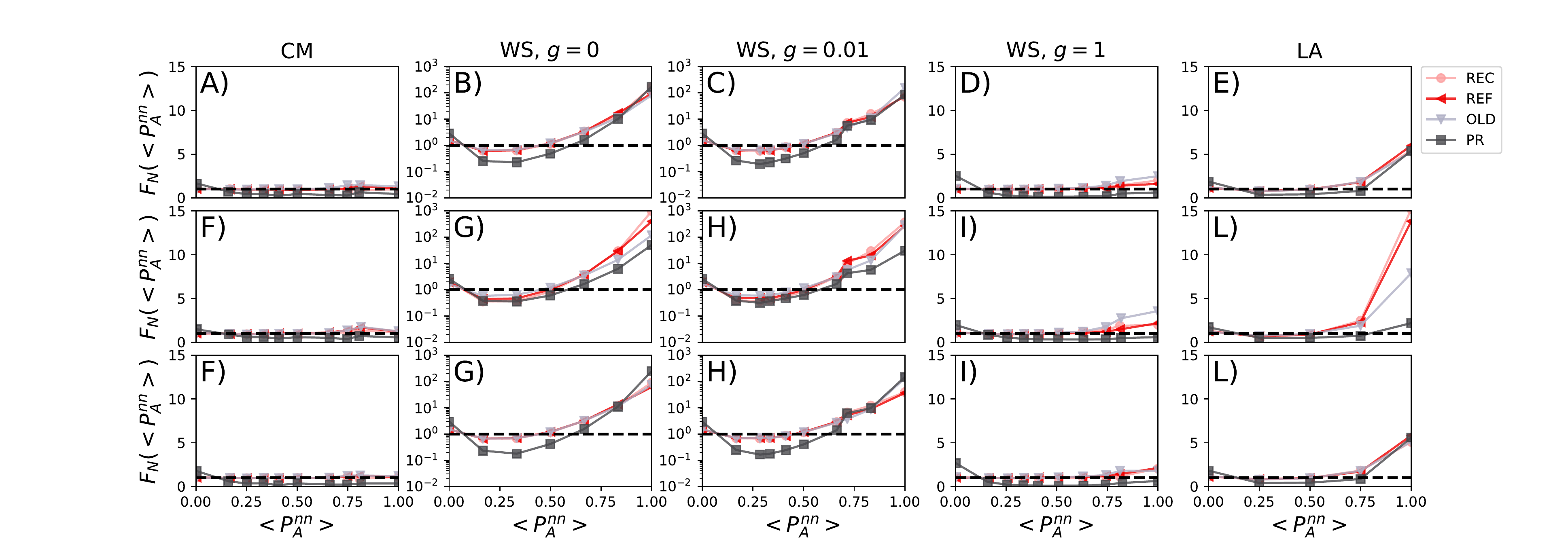}
\caption{Opinions of friends. We show the distribution of the fraction of friends (nearest neighbors, nn) of $i$ with the same opinion $A$ at $t=500$. We normalize the $y$-axis by dividing for the same quantity computed at $t=0$. We show the results for starting conditions $P_{\text{A}}(0)=0.2$ and $P_{\text{B}}(0) = 0.8$. In the first row, we show the results for $Q=20$, in the second for $Q=5$, in the third for $Q=50$. Each column described the results for a particular network. Each plot is the average of $10^2$ independent simulations.}
\label{fig:SI04}
\end{figure}

In Figure~\ref{fig:SI04} we show the same plot considering however a $20-80$ initial conditions. The results are in line with what discussed above and in the main text. In particular, topological and spatial correlation induce the formation of echo chambers around the subordinate opinion. Also in this case these effects are stronger for REC and REF filtering mechanism.

\begin{figure}[ht]
\centering
\includegraphics[width=\linewidth]{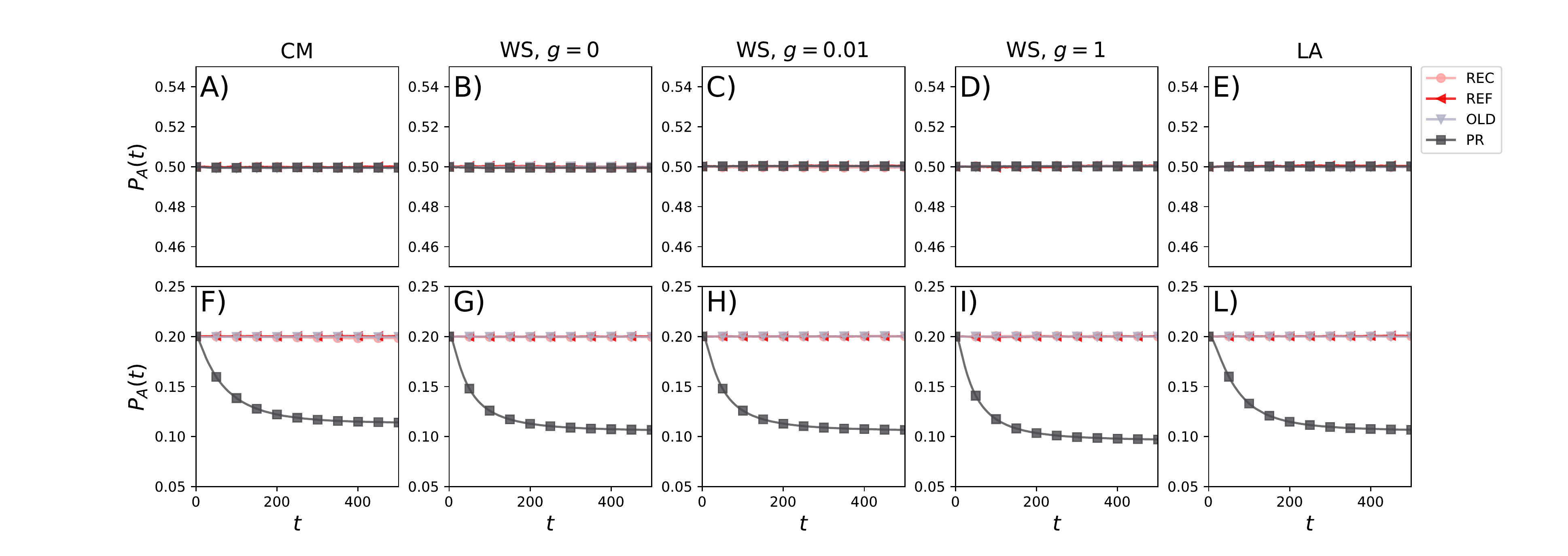}
\caption{Evolution of group opinion in the various network models. The prevalence $P_{\text{A}}(t)$ of opinion $A$ over time for starting from $P_{\text{A}}(0)=0.5$ and $P_{\text{B}}(0) = 0.5$ (first row) and $P_{\text{A}}(0)=0.2$ and $P_{\text{B}}(0) = 0.8$ (second row). Each column describes the results for one of the networks.  In each plot, we show results for the four sorting algorithms. Each plot is the average of $10^2$ independent simulations and to improve the visualization we are showing the data points every $50$ time steps. In all scenarios we set $Q=5$.}
\label{fig:SI06}
\end{figure}

In Figure~\ref{fig:SI06} we show the behavior of $P_A$ as function of time for $Q=5$. As clear from the figure, the behavior of the global share of the opinion is qualitatively similar to the other values of $Q$. For unbalanced initial conditions the semantic filtering is still the only sorting mechanism that bring to a reduction of the subordinate opinion. High clustering, spatial correlations, and heterogeneity in the degree distribution hamper the reduction respect to WS graphs with $g=1$ however to a less extent than for $Q=20$.

\subsection{$1-99$ and $10-90$ initial conditions}

\begin{figure}[ht]
\centering
\includegraphics[width=\linewidth]{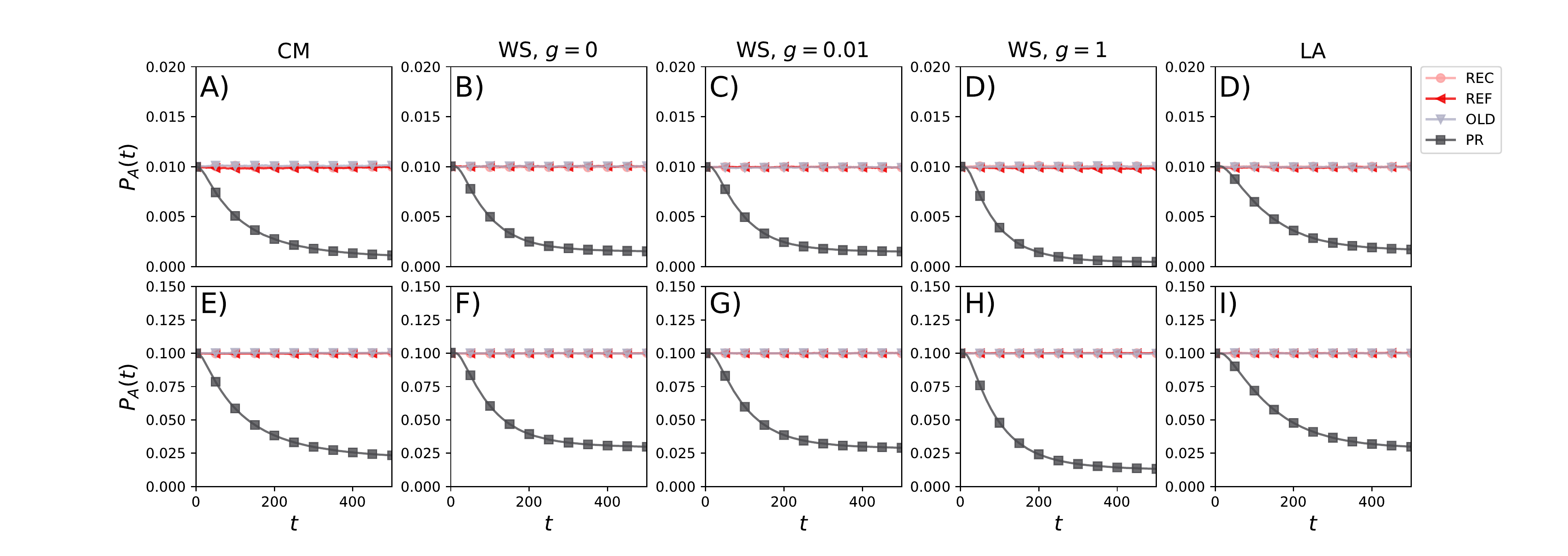}
\caption{Evolution of group opinion in the various network models. The prevalence $P_{\text{A}}(t)$ of opinion $A$ over time for starting from $P_{\text{A}}(0)=0.01$ and $P_{\text{B}}(0) = 0.99$ (first row) and $P_{\text{A}}(0)=0.1$ and $P_{\text{B}}(0) = 0.9$ (second row). Each column describes the results for one of the networks.  In each plot, we show results for the four sorting algorithms. Each plot is the average of $10^2$ independent simulations and to improve the visualization we are showing the data points every $50$ time steps. In all scenarios we set $Q=20$.}
\label{fig:SI05}
\end{figure}

We now consider the behavior of $P_A(t)$ for different unbalanced conditions. In Figure~\ref{fig:SI05} we show (first row) the case of a quite unbalanced starting scenario: $P_A(0)=0.01$ and $P_B(0)=0.99$. As the opinion model we consider is not based on a deterministic majority rule, the subordinate opinion is able to survive. Also in this case, the PR mechanisms induces a reduction which is however hampered by correlations and heterogenous connectivity patterns as shown in the main text for $20-80$ case. In the second row we show the case in which $P_A(0)=0.1$ and $P_B(0)=0.90$ which provides the same qualitatively results. 

\end{document}